\documentclass[journal]{IEEEtran}

\usepackage{cite}

\usepackage{amsmath}
\usepackage{hyperref}
\usepackage{amssymb}
\usepackage[pdftex]{graphicx}
\usepackage{float}
\newfloat{listi}{thp}{lop}
\floatname{listi}{Listing}
\usepackage[short]{optidef}
\usepackage{subcaption}
\usepackage[table,dvipsnames]{xcolor}
\usepackage[nolist]{acronym}
\usepackage[]{nomencl}
\usepackage{soul} 
\sethlcolor{Apricot} 

\RequirePackage{ifthen}
\renewcommand{\nomgroup}[1]{%
	\ifthenelse{\equal{#1}{S}}{\vspace{8pt}
		\item[\textit{Sets}]}{%
		\ifthenelse{\equal{#1}{I}}{\vspace{8pt}
			\item[\textit{Indices}]}{%
			\ifthenelse{\equal{#1}{P}}{\vspace{8pt} \item[\textit{Parameters}]}{%
				\ifthenelse{\equal{#1}{F}}{\vspace{0pt}
					\item[\textit{Functions}]}{%
					\ifthenelse{\equal{#1}{V}}{\vspace{8pt} \item[\textit{Variables}]}{}}}}}}

\begin{document}
%
\title{\texttt{PowerSimulations.jl} - A Power Systems operations simulation Library}
%
%
%

\author{Jose Daniel Lara,~\IEEEmembership{Senior Member,~IEEE,} Clayton Barrows,~\IEEEmembership{Senior Member,~IEEE,} Daniel Thom, Sourabh Dalvi, Duncan S. Callaway,~\IEEEmembership{Member,~IEEE,} and Dheepak Krishnamurthy,~\IEEEmembership{Member,~IEEE}
\thanks{This work was authored [in part] by the National Renewable Energy Laboratory, operated by Alliance for Sustainable Energy, LLC, for the U.S. Department of Energy (DOE) under Contract No. DE-AC36-08GO28308. The views expressed in the article do not necessarily represent the views of the DOE or the U.S. Government. The U.S. Government retains and the publisher, by accepting the article for publication, acknowledges that the U.S. Government retains a nonexclusive, paid-up, irrevocable, worldwide license to publish or reproduce the published form of this work, or allow others to do so, for U.S. Government purposes.}
\thanks{Corresponding author. jdlara@berkeley.edu}
\thanks{J.D.~Lara and D.S.~Callaway are with the Energy and Resources Group at the University of California Berkeley.}
\thanks{C.~Barrows, S.~Dalvi and D.~Thom are with National Renewable Energy Laboratory, Golden, CO, 80401}}

%
%

\markboth{IEEE Transactions in Power Systems}%
{Lara \MakeLowercase{\textit{et al.}}: PowerSimulations.jl- Power Systems operations simulation Library}
%



\maketitle

\begin{abstract}
\texttt{PowerSimulations.jl} is a Julia-based BSD-licensed power system operations simulation tool developed as a flexible and open source software for quasi-static power systems simulations including Production Cost Models. PowerSimulations.jl tackles the issues of developing a simulation model in a modular way providing tools for the formulation of decision models and emulation models that can be solved independently or in an interconnected fashion. This paper discusses the software implementation of PowerSimulations.jl as a template for the development and implementation of operation simulators, providing solutions to commonly encountered issues like time series read/write and results sharing between models. The paper includes a publicly-available validation of classical operations simulations as well as examples of the advanced features of the software.
\end{abstract}


%
\IEEEpeerreviewmaketitle
\begin{acronym}[SIIP]\itemsep0.01pt
\acro{ACE}{Area Control Error}
\acro{AD}{Automatic Differentiation}
\acro{AGC}{Automatic Generation Control}
\acro{AML}{Algebraic Modeling Language}
\acro{AS}{Ancillary Services}
\acro{CCO}{Chance Constrained Optimization}
\acro{CDF}{Cumulative Probability Distribution Function}
\acro{CHP}{Combined Heat and Power}
\acro{CIM}{Common Information Model}
\acro{CV@R}{Conditional Value-at-Risk}
\acro{DAUC}{Day-ahead Unit Commitment}
\acro{DER}{Distributed Energy Resource}
\acro{DG}{Distributed Generator}
\acro{DSM}{Demand Side Management}
\acro{ED}{Economic Dispatch}
\acro{EMS}{Energy Management System}
\acro{ERCOT}{Electric Reliability Council of Texas}
\acro{ESS}{Energy Storage System}
\acro{FRR}{Frequency Regulation Reserve}
\acro{GHG}{Green-House Gas}
\acro{HAUC}{Hour-ahead Unit Commitment}
\acro{IEC}{International Electrotechnical Commission}
\acro{ISO}{Independent System Operator}
\acro{LFC}{Load Frequency Control}
\acro{LP}{Linear Programming}
\acro{LQP}{Linear Quadratic Problem}
\acro{MAS}{Multi-Agent Systems}
\acro{MC}{Monte Carlo}
\acro{MILP}{Mixed-Integer Linear Programming}
\acro{MIQP}{Mixed-Integer Quadratic Programming}
\acro{MINLP}{Mixed-Integer Nonlinear Programming}
\acro{MPEC}{Mathematical Program with Equilibrium Constraints}
\acro{MP}{Mathematical Program}
\acro{MPC}{Model Predictive Control}
\acro{MT}{Micro-Turbine}
\acro{NLP}{Nonlinear Programming}
\acro{NWP}{Numerical Weather Prediction}
\acro{OPF}{Optimal Power Flow}
\acro{PCM}{Production Cost Model}
\acro{PEC}{Power Electronics Converter}
\acro{PF}{Power Flow}
\acro{PDF}{Probability Density Function}
\acro{PMF}{Probability Mass Function}
\acro{PV}{Photovoltaic}
\acro{PWL}{Piece-wise Linear}
\acro{RE}{Renewable Energy}
\acro{RO}{Robust Optimization}
\acro{RUC}{Robust Unit Commitment}
\acro{PI}{Proportional Integral}
\acro{SACE}{Smooth Area Control Error}
\acro{SO}{Stochastic Optimization}
\acro{SCIG}{Squirrel-cage Induction Generator}
\acro{SoC}{State-of-Charge}
\acro{SP}{Stochastic Programming}
\acro{SDDP}{Stochastic Dual Dynamic Programming}
\acro{SCED}{Security Constrained Economic Dispatch}
\acro{SUC}{Stochastic Unit Commitment}
\acro{UC}{Unit Commitment}
\acro{V@R}{Value-at-Risk}
\acro{VRE}{Variable Renewable Energy}
\acro{VSC}{Voltage Source Converter}
\acro{WT}{Wind Turbine}
\end{acronym}
\section{Introduction}
%
%
%
%
\IEEEPARstart{W}ITH the integration of large amounts of \ac{VRE} in electric grids, there is a need to advance the flexibility and robustness of industry practices. Optimization-based models have become an essential tool in power systems operations to efficiently and reliably plan and schedule grid operations \cite{KROPOSKI2017}. Assessing the effects of \ac{VRE} expansion and demonstrating the value of novel approaches over conventional operation strategies relies on computer simulations as a means of testing. The outcomes of such simulations are of great use to diverse stakeholders---including planners, analysts, and regulators--- who depend on operational model simulations for analyses that range from the cost of supply estimation to expansion plan feasibility. In many cases, the outcomes of these analyses are relevant to the public, the scientific community, or both. Given this breadth of applications for power systems simulations, transparency and reproducibility \cite{pfenninger2018opening} are imperative as they allow the verification of results \cite{decarolis_case_2012}, which can be accomplished with rigorous scientific computing practices. Scientific computing practices promote the application of principles such as reproducibility, transparency, and accuracy to experiments carried out using computer simulations \cite{LARA2020106680}. There is large and growing relevance of scientific computing for energy simulations that should provide adequate access to the models and the data used.

One of the key innovations arising from systems operations modeling has been the inclusion of sub-hourly effects. Recent contributions show the relevance of intra-period coordination is operations assessment \cite{plexos, wang2016quantifying}. The main driver to assess the intra-temporality research questions is software that simulates system operations with sequentially coupled optimization problems. Handling large scale and long range simulations, however, has largely been limited to commercial software tools like PLEXOS, PROMOD or GE-MAPS. Given the complex, data-intensive nature of an operations simulation and the general lack of access to source code and data, many of the underlying assumptions remain hidden to external observers \cite{decarolis_case_2012}.

Models used to simulate power system operations usually feature significant simplifications of the system's physics to formulate optimization problems that focus on power system decision-making such as commitment of units, dispatch or reserve allocation. Hence, a transparent framework requires that the simulation experiment properly account for all the assumptions that go into the models' simplifications, particularly when using the simulations to produce scientific or policy statements.  However, commercial simulation software models are intellectual property not available to the public and their cost can be a significant barrier to reproducibility. When commercial simulators are not available, power system operations researchers have to develop \textit{ad-hoc} heuristics and solutions to combine optimization and modeling software from multiple vendors. The resulting lack of coherence across platforms and reliance on bespoke frameworks to execute experiments makes simulation evaluation, adoption, modification, and expansion costly, and in many cases it is impossible to reproduce \cite{LARA2020106680}.

In recent years, open-source operations simulation software has become available for research purposes. One example of software that to date is actively maintained is the Python library \texttt{PyPSA} \cite{PyPSA} which focuses on expansion planning models and enables users to assess the plans that result from simulating operations. Another is \texttt{FESTIV} \cite{ela_flexible_2011}, a primarily Matlab-based software that relies on a mixture of GAMS and Matlab for the study of the impact of \ac{VRE} in system operations at high levels of granularity. Although the code for FESTIV is publicly available, it requires commercial software, while model modification requires changing the source code. Finally, the Julia package \texttt{PowerModels.jl} \cite{coffrin2018powermodels} focuses on the formulation and solution of \ac{OPF} problems. Although \texttt{PowerModels.jl} is capable of handling realistically-sized systems and is openly sourced, it cannot inherently be used to perform long-range simulation, i.e., representing months to years of operation, or solve sequences of operational problems.

\subsection{Motivations}

Power systems research has benefited from having open source analysis and simulations tools. However, existing operations simulations tools have focused on single-period problems (e.g. load flow and contingency analysis) or  multi-period problems like \ac{UC} or \ac{OPF}, thus avoiding the complexities of inter-problem information flow.  Despite the widespread usage of \ac{PCM} in research, and industry decision-making, the definitions and algorithmic structure that enables \ac{PCM} simulations is poorly represented in literature. The lack of availability of operations simulation software drives the lack of reproducible research in this area.

Furthermore, given that the alternatives to developing a large scale operations simulation results in complex software projects or depend on commercial tools, the underlying optimization models used in the simulation tend to be rigidly defined. The model rigidity can constrain the questions that researchers and analysts can examine, resulting in \emph{model-limited choice} \cite{dorris2021choice, pechman1993model}. The lack of flexibility characterizing the models and operational sequences influences the type of simulations that can be conducted and in turn the scientific inquiries. Model-limited choices can stem from two factors: 1) structural exclusion of certain forms of simulation and analysis; and 2) formulation limitations due to restrictions in the underlying models or whether data is available.

The objective of \texttt{PowerSimulations.jl} is two-fold: 1) enable a scientific approach to the simulation of large scale power system operations; and 2) reduce model-limited choice when framing operations simulation experiments. \texttt{PowerSimulations.jl} also aims to provide the necessary utilities to develop simulations at a scale and scope on par with commercial tools.

\subsection{Structure and scope}
This paper focuses first on introducing the definition of a \emph{operations simulation} along with other terminology required to formulate a \emph{operations simulation model}. We focus on the technical aspects implemented in \texttt{PowerSimulations.jl} that enable the scalability and flexibility of \emph{operations simulations}. Throughout the paper we introduce definitions to the properties of operations simulations that are commonly used but rarely discussed in the simulation literature. The contributions of this paper are:

\begin{itemize}
    \item The formal definition of an operations simulation that employs sequential solves of multiple optimization problems.
    \item A systematic method to formulate operation optimization models for the purpose of simulation incorporating devices, network and services that span multiple time resolutions.
	\item The description of methods and software implementation requirements to support long range simulations.
	\item A framework that enables developers and researchers to scientifically reproduce an operations simulation.
\end{itemize}

The rest of the paper is organized as follows: Section \ref{sec::defs} discusses the definitions of an operations simulation and provides a robust description of its atomic components. Section \ref{sec::software} showcases the software implementation of an operations simulation package, with consideration to two main components: namely, model building and execution. Section IV examines the validation and example uses of \texttt{PowerSimulations.jl} to publicly available data sets. Finally, Section V discusses conclusions and future work.

\section{Defining an operations simulation} \label{sec::defs}

\subsection{Simulation vs Optimization}


Power systems operations is a rich field for the study and development of \emph{optimization models} to find the solution of \emph{operation problems}. Given its importance, and the complexity and scale of the electric grid, there exists an abundance of literature on approaches that efficiently formulate and solve \emph{optimization models} with applications to power systems \cite{panciatici2014advanced}. For instance, \ac{UC} is an operation problem typically formulated as a \ac{MILP} optimization model and has been the focus of countless formulations and algorithmic improvements. Also, the \ac{OPF} problem has received significant contributions with the development of relaxed \emph{optimization model} formulations that exploit recent advances in convex optimization algorithms.

On the other hand, the objective of an \emph{operations simulation} is to find solutions to a \emph{sequence} of \emph{operation problems} in a way that resembles the procedures followed by operators. For instance, an \emph{operations simulation} seeks to replicate the day-ahead \ac{UC}, hour-ahead \ac{UC} and real-time \ac{ED} market clearing processes. The results of an operations simulation could be used to assess the generation fleet fuel costs during a year. The evolution from ``load-duration" curves to assess systems costs to the use of optimization has enabled increased insight into the effects \ac{VRE} in the system operation \cite{plexos}. However, as the complexity of the analysis grows to incorporate new technologies or bring a more detailed representation of the system's operation into the simulation, the scope of \ac{PCM} has grown beyond its original application. Hence, a simulation requires the specification of one or more \emph{operation problems} and \emph{optimization models} used to represent and solve each problem. Each \emph{optimization model} is an \emph{atomic} component of a simulation, while the formulation of an \emph{operation problem} can differ depending on the application, jurisdiction, or phenomena of interest.

\subsection{Operation Problem Definitions}

\begin{figure}[t]
    \centering
    \includegraphics[width=\columnwidth]{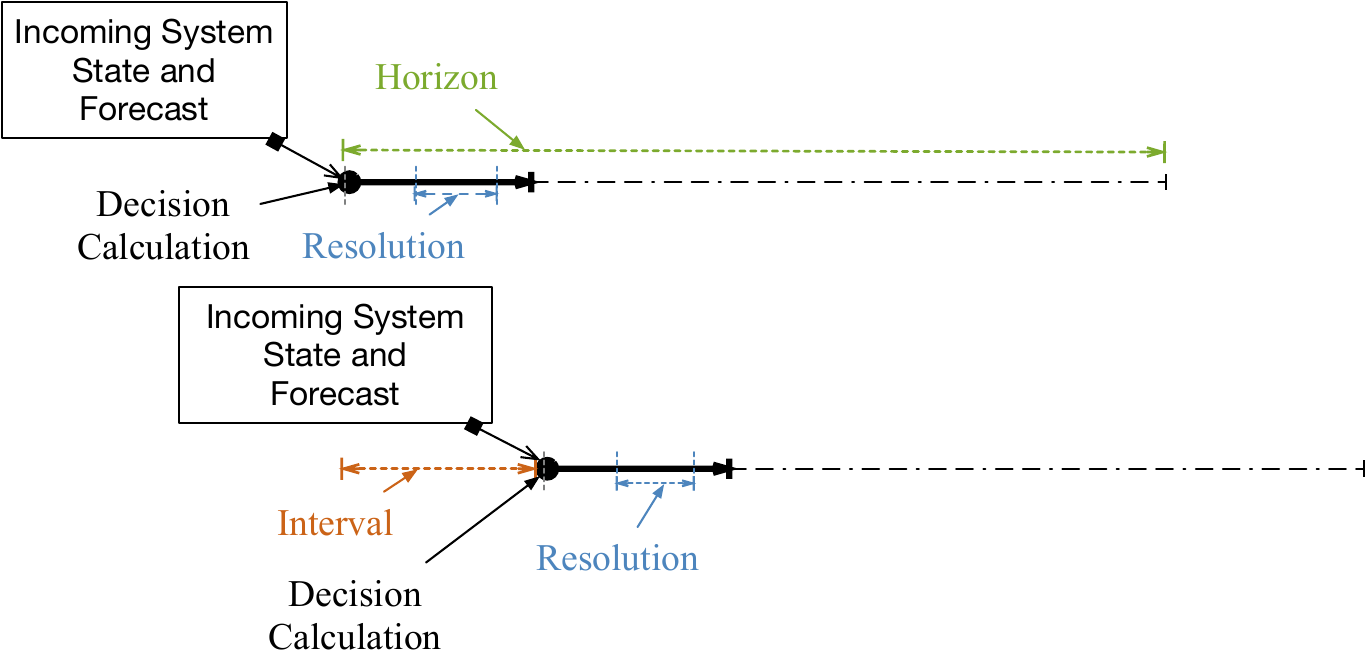}
    \caption{Sequential properties of a decision model. Each calculation step is taken at a particular interval and employs information about the state of the system and a forecast to estimate future estates.}
    \label{fig:decision_defs}
\end{figure}

\begin{figure}[t]
    \centering
    \includegraphics[width=0.7\columnwidth]{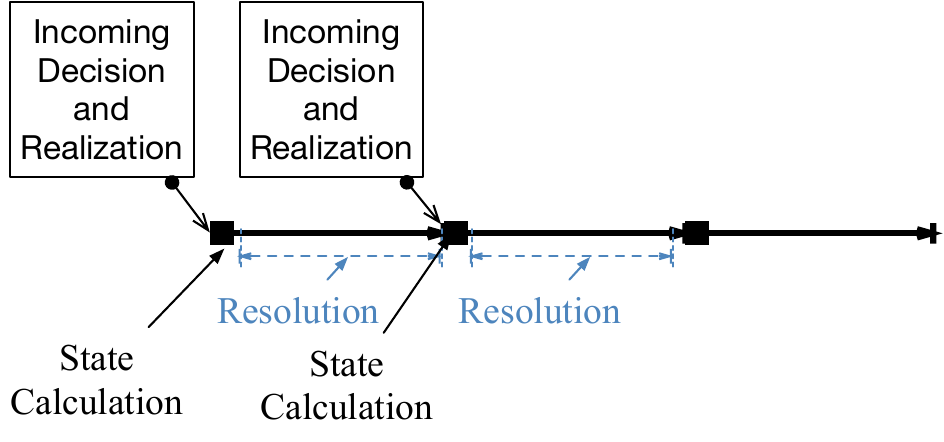}
    \caption{Sequential properties of an emulation model. At each calculation step, the model uses the incoming decisions and the realization of the uncertain quantities such that the state values at a time $t$ are independent of future system states.}
    \label{fig:emulator_defs}
\end{figure}

We begin by defining a \textit{decision} problem and an \textit{emulation} problem, two types of operation problems used in the course of an operations simulation. Both types of operation problems are discretized to a specific \emph{resolution} $\Delta t$ depending on the requirements and available input data.
\begin{itemize}
    \item {\textbf{Decision Problem:} A decision problem calculates the desired system operation based on forecasts of uncertain inputs and information about the state of the system. The output of a decision problem represents the policies used to drive the set-points of the system's devices, like generators or switches, and depends on the purpose of the problem. The decision problems employ the forecasts to make decisions for the discrete time-steps over a \emph{horizon}. The model of a decision problems updates the decision values at an \emph{interval} which is larger than its \emph{resolution}, Fig. \ref{fig:decision_defs} depicts these definitions. For instance in day-ahead operations the common practice is to use a \ac{UC} problem with a 24-hour \emph{interval}, 1-hour \emph{resolution} and over a 48-hour \emph{horizon}. The decisions calculated in the portion of the horizon after the end of the interval (dash line in Fig. \ref{fig:decision_defs}) are not implemented and are used as a way to handle inter-temporal effects of the uncertain quantities\footnote{Some operators call that segment of the horizon the "additional look-ahead".}.} Decisions in power systems are calculated in a \emph{sequential} or \emph{staged} manner since decisions are continually refined as information about the operating period updates.
    \item {\textbf{Emulation Problem:} An emulation problem is used to mimic the system's behavior subject to an incoming decision and the realization of a forecasted inputs. The solution of the emulator produces outputs representative of the system performance when operating subject the policies resulting from the decision models. The emulator model is ``myopic" and executed along a single timeline, as shown in Fig. \ref{fig:emulator_defs}, since its results are not affected by future system states. For instance, an \ac{AGC} model can be used as an emulator problem to evaluate the deployment of reserves. Alternatively, an AC power flow can be used to assess the resulting node voltages after a dispatch decision. The emulator also plays a critical role in simulations with inter-temporal constraints since it can better represent the initial conditions in models that require initial condition values like those featuring ramp constraints or storage devices.
}
\end{itemize}

Conceptually, the relationship between the decision problem and the emulator problem is akin to the control-plant model commonly used in the field of automatic control. The representation of the system state through an emulation problem is not common practice in \ac{PCM}. This is due in part to the fact that most commercial \ac{PCM} tools only support a one- or two-problem simulation framework. Extending an operations simulation to use multiple problems is significantly more challenging.

\subsection{Model of an operations simulation}
\begin{figure}[t]
    \centering
    \includegraphics[width=\columnwidth]{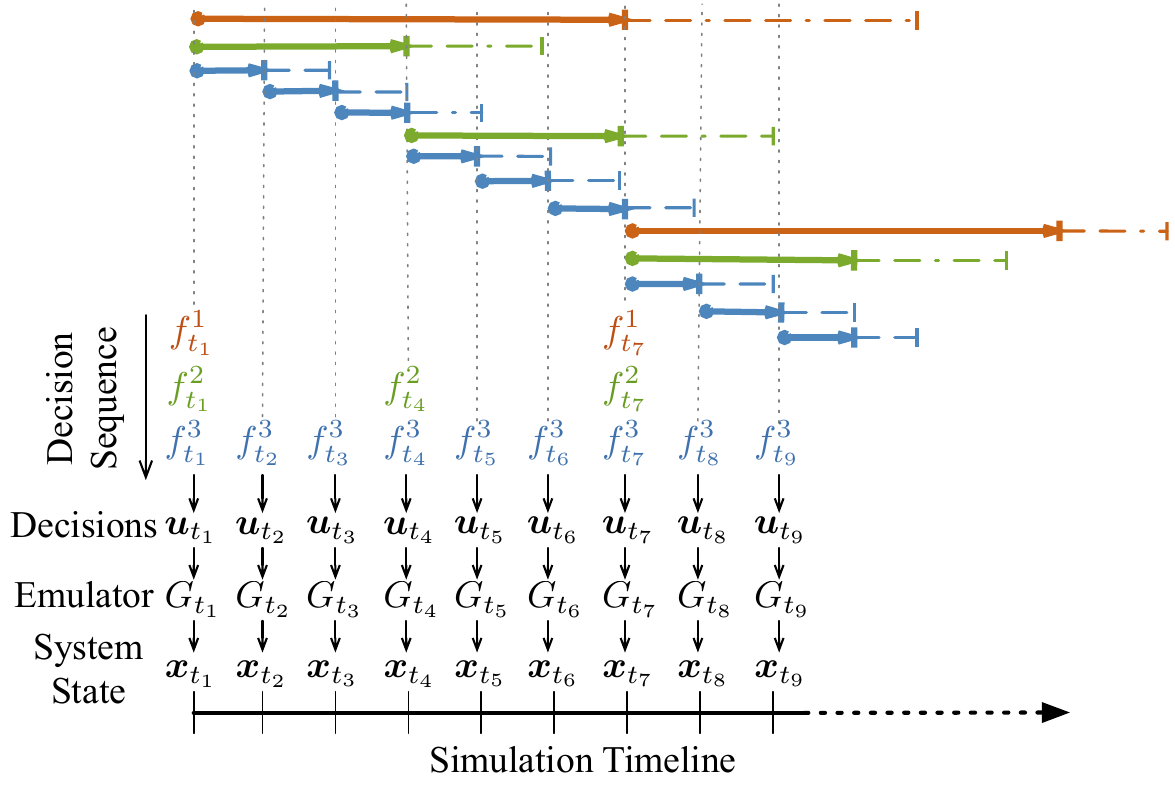}
    \caption{operations simulation chronological sequence of calculations.}
    \label{fig:sequence_example1}
\end{figure}
Based on the previous definitions of decision and emulation problems, it is possible to formulate an operations simulation using a discrete time model as follows:
\begin{align}
    \boldsymbol{u}_{t} \! = \! F_t(\boldsymbol{x}_{t-1}, \boldsymbol{u}_{t-1},\boldsymbol{\rho}_t, \Phi|t), &\quad \boldsymbol{u}_{t_0} \! = \! \boldsymbol{u}_0 \label{eq:sim1}\\
     G_t(\boldsymbol{x}_{t}, \boldsymbol{x}_{t-1}, \boldsymbol{u}_t, \boldsymbol{\psi}_t) \! = \! 0 , &\quad \boldsymbol{x}_{t_0} \! = \! \boldsymbol{x}_0 \label{eq:sim2}
\end{align}
\nomenclature[P]{$\Phi|t$}{Forecast data issued for time $t$}
\nomenclature[P]{$\boldsymbol{u}_{t}$}{Decision value at $t$}
\nomenclature[P]{$\boldsymbol{u}_{t}$}{System state at $t$}
\noindent where $\boldsymbol{u}_{t}$ represents the operation decisions conditional on a forecast issued for time $t$, $\Phi|t$ over a horizon $\mathcal{H}$ and using parameters $\boldsymbol{\rho}$. The decision state $\boldsymbol{u}_t $ is calculated over a horizon $\mathcal{H}$ such that $\boldsymbol{u}_t = \left \{ u_{h|t} | ~h \in \mathcal{H} \right \}$  where $u_{t|h}$ represents a decision taken at time $t$ for time-step $h$. The decision variables are updated by function $F_t$ that represents the sequential solution of decision problems. Function $G_t$ is the emulator problem that updates the system state $\boldsymbol{x}_{t}$ given the decision $\boldsymbol{u}_{t}$ and the realized inputs $\boldsymbol{\psi}_t$.

Given \eqref{eq:sim1}-\eqref{eq:sim2}, a simulation can be set up as follows: given an initial condition for a decision and system state, advance in time $t$ from one point to the next considering a discrete timeline $\{t_0,t_1,…,t_n,…,T\}$. A simulation requires a stepping procedure that finds the solution in time $t_{n+1}$ provided the values of the variables at $\{t | t_0 \le t < t_{n+1}\}$.

At each time-step $t$ the sequential solution of decision problems $F_t$ is formulated as a composition of functions over a set $\mathcal{K}_t$:
\begin{equation}
    F_t \! \coloneqq \! \!f^k \circ\! f^{k-1} \! \circ \dotsm \circ\! f^1 (\boldsymbol{x}_{t-1}, \boldsymbol{u}_{t-1}, \boldsymbol{\rho}_t, \Phi|t)
\end{equation}
\noindent each $f^k$ corresponds to solving an optimization model to update the decision states based on the previous decisions $ \boldsymbol{u}_{t-1}$, the system state $\boldsymbol{x}_{t-1}$ and the available forecast $\Phi^k|t$. Figure \ref{fig:sequence_example1} shows an example of the stepping process in a simulation, where the sequence $F_t$ can have between 1 and 3 function evaluations to arrive to the decision value $\boldsymbol{u}_{t}$. Before the next set of decisions are updated the emulation problem is evaluated.

Each function $f^k$ is an optimization model of a sub-set of the decisions $\boldsymbol{u}^k_t$ using a sub-set of the forecast data $\Phi^k|t$ as follows:
\begin{subequations}
\begin{align}
f^k(\cdot) =\min_{\boldsymbol{u}^k_t} & \quad C_{f_k}(\boldsymbol{u}^k_t)\\
 \text{s.t.}
& \quad H^{D}_{f_k}\left(\boldsymbol{u}_t, \boldsymbol{u}_{t-1},\boldsymbol{x}_{t-1}, \boldsymbol{\rho}_t, \Phi^k|t \right) \le 0 &\quad\\
& \quad H^{B}_{f_k}\left(\boldsymbol{u}_t, \boldsymbol{u}_{t-1},\boldsymbol{x}_{t-1}, \boldsymbol{\rho}_t, \Phi^k|t \right) \le 0 &\quad\\
& \quad H^{N}_{f_k}\left(\boldsymbol{u}_t, \boldsymbol{u}_{t-1},\boldsymbol{x}_{t-1}, \boldsymbol{\rho}_t, \Phi^k|t \right) = 0 &\quad\\
& \quad H^{S}_{f_k}\left(\boldsymbol{u}_t, \boldsymbol{u}_{t-1},\boldsymbol{x}_{t-1}, \boldsymbol{\rho}_t, \Phi^k|t \right) \le 0 &\quad\\
& \quad H^{F}_{f_k}\left(\boldsymbol{u}_t, \boldsymbol{u}_{t-1},\boldsymbol{x}_{t-1}, \boldsymbol{\rho}_t, \Phi^k|t \right) \le 0 &\quad
\end{align}
\label{eq:op_model}
\end{subequations}
\noindent $C_{f_k}$ is the objective function of the model. The constraint function $H^{D}_{f_k}$ corresponds to the models for the injection devices like generation, loads, and storage. Constraints $H^{B}_{f_k}$ correspond to the models of the branches like Lines and HVDC, while $H^{N}_{f_k}$ corresponds to the model of the network like CopperPlate, DC powerflow, or AC powerflow. $H^{S}_{f_k}$ correspond to the models of the services like reserves or transfers. Finally, $H^{F}_{f_k}$ are introduced as the \emph{feedforward} constraints.

\emph{Feedforwards} constraints are used to represent the relationships that the decision variables $\boldsymbol{u}_{t}$ have between the models. For instance,  a \emph{feedforward} is commonly used to pass the values of the on/off decision variables from a \ac{UC} problem into an \ac{ED} problem. Other examples of \emph{feedforward} constraints include upper-bounds, lower-bounds, and end-of-horizon state of charge targets among others. These constraints are not commonly formalized in commercial or open source simulation models despite the fact that their implementation can affect the model's results.

\begin{figure}[t]
    \centering
    \includegraphics[width=\columnwidth]{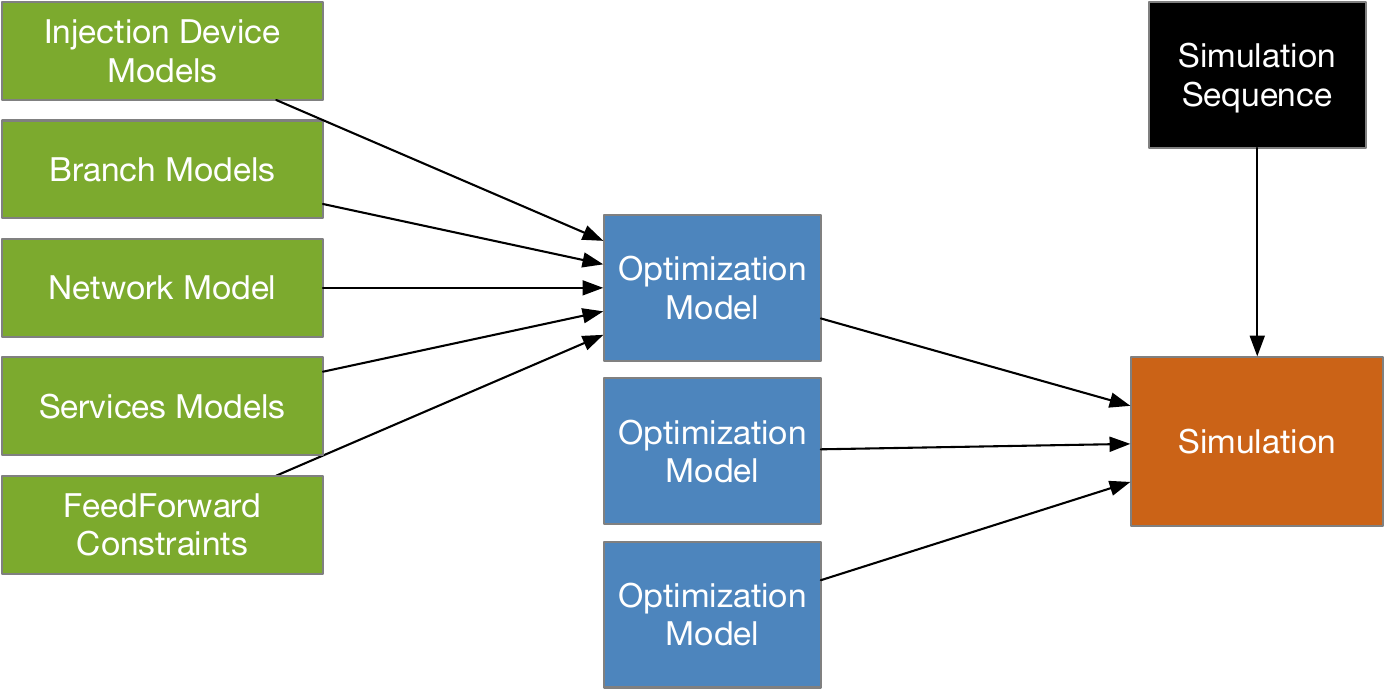}
    \caption{Model relationship graph of a generic operations simulation.}
    \label{fig:sim_data_relationships}
\end{figure}

\section{Software Structure and Description} \label{sec::software}

An implementation of an operations simulation software needs to consider two large workflows: 1) building the optimization model; and 2) handling the sequencing and book-keeping of the optimization model solutions.

From a software perspective, the challenge of implementing a scalable and flexible simulation framework lies in handling the required data structures. As described in \cite{LARA2020106680}, building and solving optimization model within a simulation requires several data translations. Traditionally, \ac{PCM} simulators employ disk writes to handle these data transactions; at each model solve the software writes a text file to disk that's later solved by the optimizer. In an optimization context, the limitations of this workflow are further discussed in detail by \cite{DunningHuchetteLubin2017}, where the authors detail the implementation of the \ac{AML} \texttt{JuMP.jl} and the comparison with text-interpreted alternatives like GAMS or AMPL.
From a simulation perspective, this workflow is inefficient since it implies that the solver garbage collects the incumbent solution that can be used as an initial guess and for models that employ \ac{LP} or \ac{MILP} solvers it implies losing matrix factorizations and other intermediate values that can accelerate the solution of the model. \texttt{PowerSimulations.jl} addresses the limitations of existing simulation workflows by exploiting the in-memory representation of the optimization model provided by \texttt{JuMP.jl} \cite{DunningHuchetteLubin2017} and the underlying abstract solver interface \texttt{MathOptInterface.jl} \cite{legat2021mathoptinterface}.

\subsection{Optimization Models Build}
\begin{listi}
\includegraphics[width=\columnwidth]{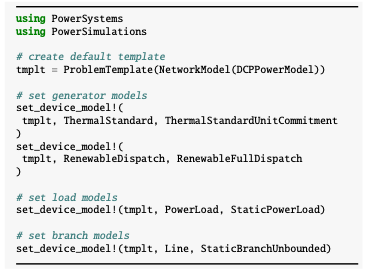}
 \caption{Definition of a problem template. High level specification of $H_{f_k}$ functions in \eqref{eq:op_model}}
 \label{lst::template}
\end{listi}
To minimize issues of model-limited choice, the optimization model building implementation follows two principles:
\begin{itemize}
    \item Flexibility: a change to a component model does not affect the model of an unrelated component
    \item Modularity: separate the individual models of the system components.
\end{itemize}

One critical consideration when building \eqref{eq:op_model} programmatically is providing sufficient flexibility for the exploration of a hypothesis while maintaining a certain level of structure such that a software solution can be implemented. In \texttt{PowerSimulations.jl}, we divide the optimization model in the sub-models described by \eqref{eq:op_model} and shown in Figure \ref{fig:sim_data_relationships}, where each simulation is comprised of optimization models formulated from component level models and the feedforwards.

The specifications of a problem are contained in a \emph{template} that determines the formulations for each component. An example of building an operation's problem template is shown in Listing \ref{lst::template}. In this example the template is based on a DC Powerflow, with the specific models for each component explicitly set. For example, \texttt{ThermalStandard} devices are modeled using the formulation \texttt{ThermalStandardUnitCommitment}. \texttt{PowerSimulations.jl} provides a comprehensive and fully documented library of formulations for diverse components in the documentation. To tackle the requirements discussed in \cite{decarolis_case_2012, pfenninger2018opening} and provide users with access to the model details and data, the optimization model objects are stored in the optimization container. This container is serialized to disk in order to guarantee that the simulation can be reproduced at a later time.

\begin{figure}[t]
    \centering
    \resizebox{\columnwidth}{!}{
    \includegraphics[width=\columnwidth]{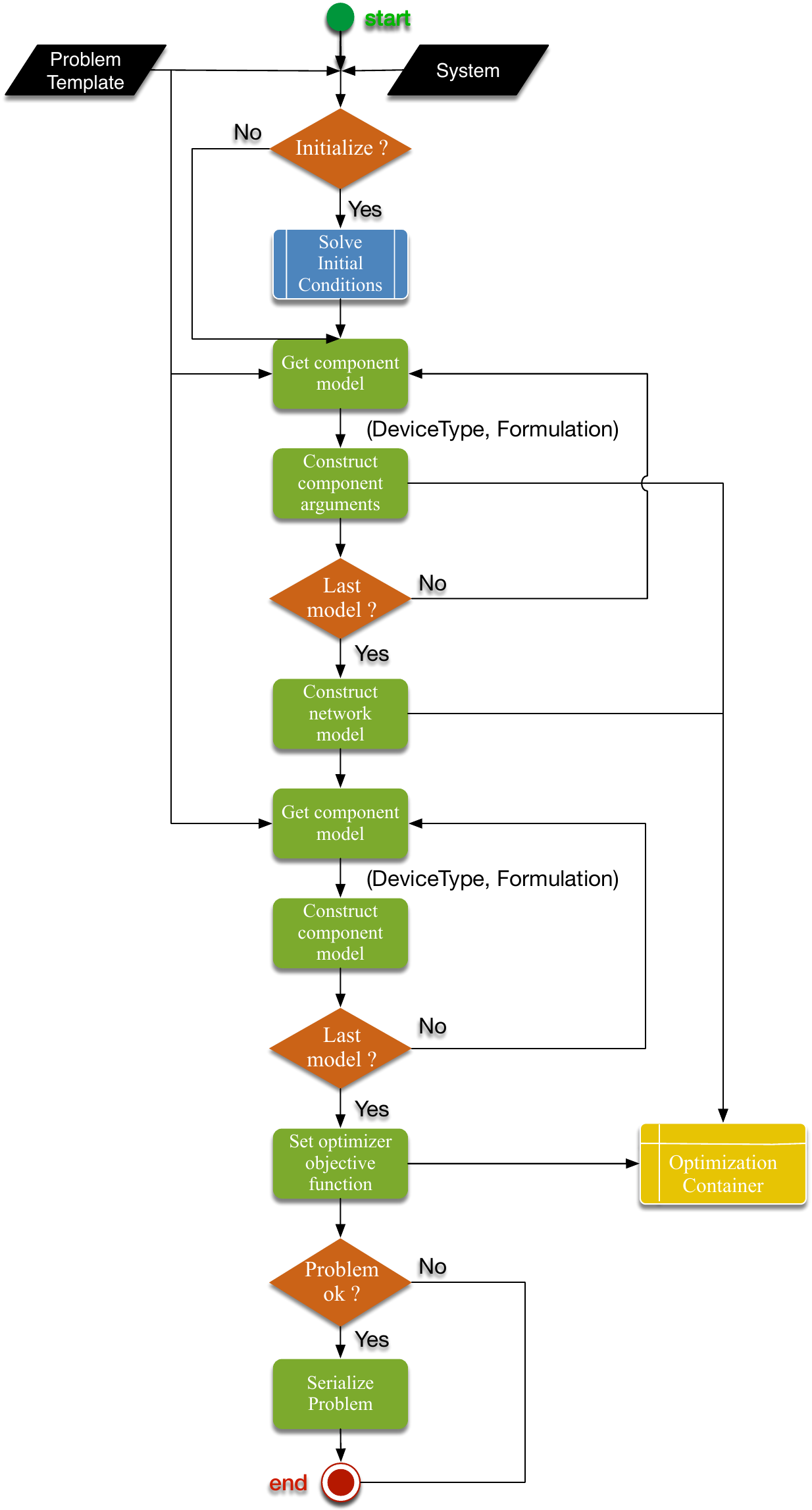}
    }
    \caption{Modular Optimization Build Flow Chart.}
    \label{fig:construction_flow}
\end{figure}

The optimization model build process is shown in Figure \ref{fig:construction_flow}. First, if a feasible initial point is not provided, \texttt{PowerSimulations.jl} builds and solves a relaxed optimization model to obtain valid values for $\boldsymbol{u}_0$ and $\boldsymbol{x}_0$ based on the template. This feature reduces the possibility of having infeasible problems at the start of the simulation due to mismatches in the incoming data.
The mathematical description for a device is implemented using Julia's multiple dispatch. The construction methods can be defined based on abstract data structures to enable code re-use and interfacing with other models \cite{bezansonJulia} which enables greater flexibility in modeling choice. \texttt{PowerSimulations.jl} builds the optimization model using the structure hierarchy defined in \texttt{PowerSystems.jl} and the data handling features \cite{lara2021powersystems}. For each component type and formulation, \texttt{PowerSimulations.jl} adds the specified arguments (i.e., variables, parameters and expressions).

\texttt{PowerSimulations.jl} implements \emph{parameters} that enable the in-memory-modification of the model instead of the common practice of using ``dummy" variables. \emph{Parameters} are implemented by keeping in memory the constraint index of the parameter and request precise model modifications without rebuilds. The limitation of this approach that the parameters can only be used to update the right-hand side of linear constraints and linear objective function coefficients. However, for most applications the inputs that change over the course of a simulation such as forecasts, feedforwards or costs can be readily implemented within this limitation.

After the arguments are built, the network formulation adds the constraints required to model the power flow through the network into the optimization model.  The modularity features in \texttt{PowerSimulations.jl} design allows an seamless integration with \texttt{PowerModels.jl} \cite{coffrin2018powermodels} which implements the network model. As a result, users can explore distinct network formulations that have been validated and tested.
Next, the device's models are constructed (i.e., adding the constraints into the model) and the objective function is set. After the process concludes, we perform a series of checks to the problem to determine if there are scaling issues or invalid values in the constraints. If the process succeeds, the model is serialized to disk for record keeping purposes.

\begin{listi}
\includegraphics[width=\columnwidth]{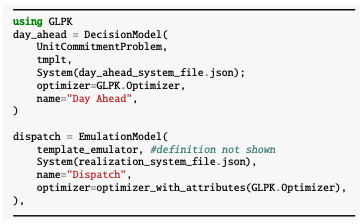}
 \caption{Definition of a \ac{UC} decision problem based on the template from Listing \ref{lst::template} and an emulation problem}
 \label{lst::decision_model}
\end{listi}


Listing \ref{lst::decision_model} shows the specification of a decision and emulation models using the template. The design allows modelers to mix and match formulations for the devices according to their needs.

\subsection{Simulation Sequence}

\begin{listi}
\includegraphics[width=\columnwidth]{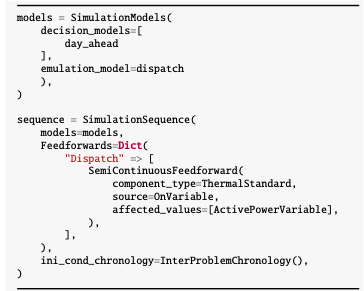}
 \caption{Definition of a SimulationSequence}
 \label{lst:sequence}
\end{listi}

A simulation also requires setting the cadence of the problems' sequential solves and specification of how information regarding $\boldsymbol{u}_t$ is exchanged between the decision and the emulator models. In \texttt{PowerSimulations.jl} the \emph{SimulationSequence} plays three roles in the specification of a simulation: (1) It performs series data validations at all times to guarantee that the intervals, resolution, and horizons in the models have the consistency required to execute a simulation; (2) it is used to set the models' \emph{feedforwards}; and 3) how the results are used to set the models' initial conditions (i.e., the \emph{initial conditions chronology}).

Listing \ref{lst:sequence} shows the example of specifying the sequence for a simulation that uses the models defined in Listing \ref{lst::decision_model}. The example sets a \emph{SemiContinousFeedforward} between the \ac{UC} problem and the emulator. This feedforward introduces the constraint:
\begin{equation}
    \text{ON}_{g,t}P^{lb}_g \le p_{g,t} \le \text{ON}_{g,t}P^{lb}_g \quad \forall g \in \mathcal{G}_{th} \ t \in \mathcal{T}
\end{equation}
\noindent where $\text{on}_{g,t}$ is a parameter in the model to input the values of $\boldsymbol{u}_t$. $p_{g,t}$ are the active power output variables of the set of thermal generators $\mathcal{G}_{th}$ at time $t$.

Listing \ref{lst:sequence} also shows the explicit handling of the initial conditions. We implement two \emph{chronologies} commonly used in simulations. The \emph{InterProblemChronology} which implies that the initial conditions are updated based on the system state and assumes that at the moment to update the decisions the model has knowledge of the state of the system.\footnote{\emph{InterProblemChronology} is similar to the ``interleaved" simulations used in some commercial simulators} The other alternative is the \emph{IntraProblemChronology} which uses the results of the decision problems as initial conditions, this chronology's assumption is that the initial conditions in the decision models are estimates resulting from previous solutions of the decision variables and taken without information regarding the system's state.

\subsection{Simulation}

The simulation is then fully specified by the models and previously defined sequence. By allowing the user to arrive to the simulation by instantiating intermediate objects, it is possible to modularize the process and facilitate debugging. Once the model inputs and sequence have been validated, the simulation can be built and executed as shown in Listing \ref{lst:sim}.

\begin{figure}[t]
    \centering
    \resizebox{\columnwidth}{!}{
    \includegraphics[width=\columnwidth]{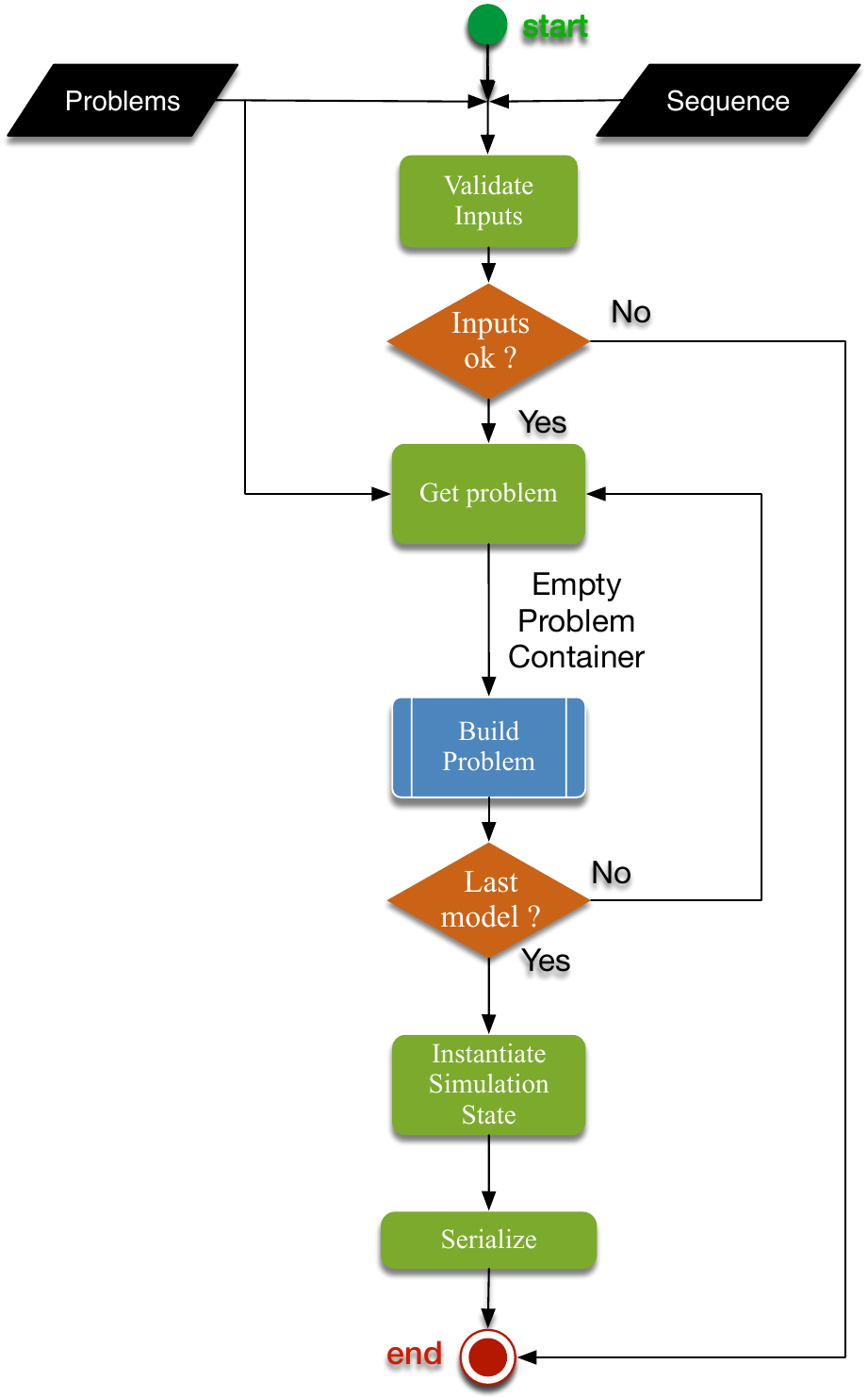}
    }
    \caption{Simulation Build Flow Chart.}
    \label{fig:sim_construction_flow}
\end{figure}

\begin{listi}
\includegraphics[width=\columnwidth]{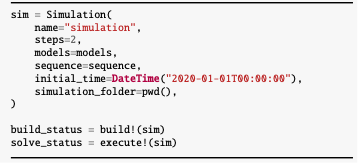}
 \caption{Definition of a Simulation and build/solve call}
 \label{lst:sim}
\end{listi}

The build process of the simulation is shown in Fig. \ref{fig:sim_construction_flow}. Once the models and sequence are validated, the optimization model corresponding to each problem is built following the process in Fig. \ref{fig:construction_flow}. Note that given the in-memory problem modification design, the simulation build process is the only time that the models are built. Once the models are built and the optimizers instantiated, the simulation state is built by pre-allocating the data structures to keep in-memory the values of $\boldsymbol{u}$ and $\boldsymbol{x}$ in-memory.


\begin{figure}[t]
    \centering
    \resizebox{\columnwidth}{!}{
    \includegraphics[width=\columnwidth]{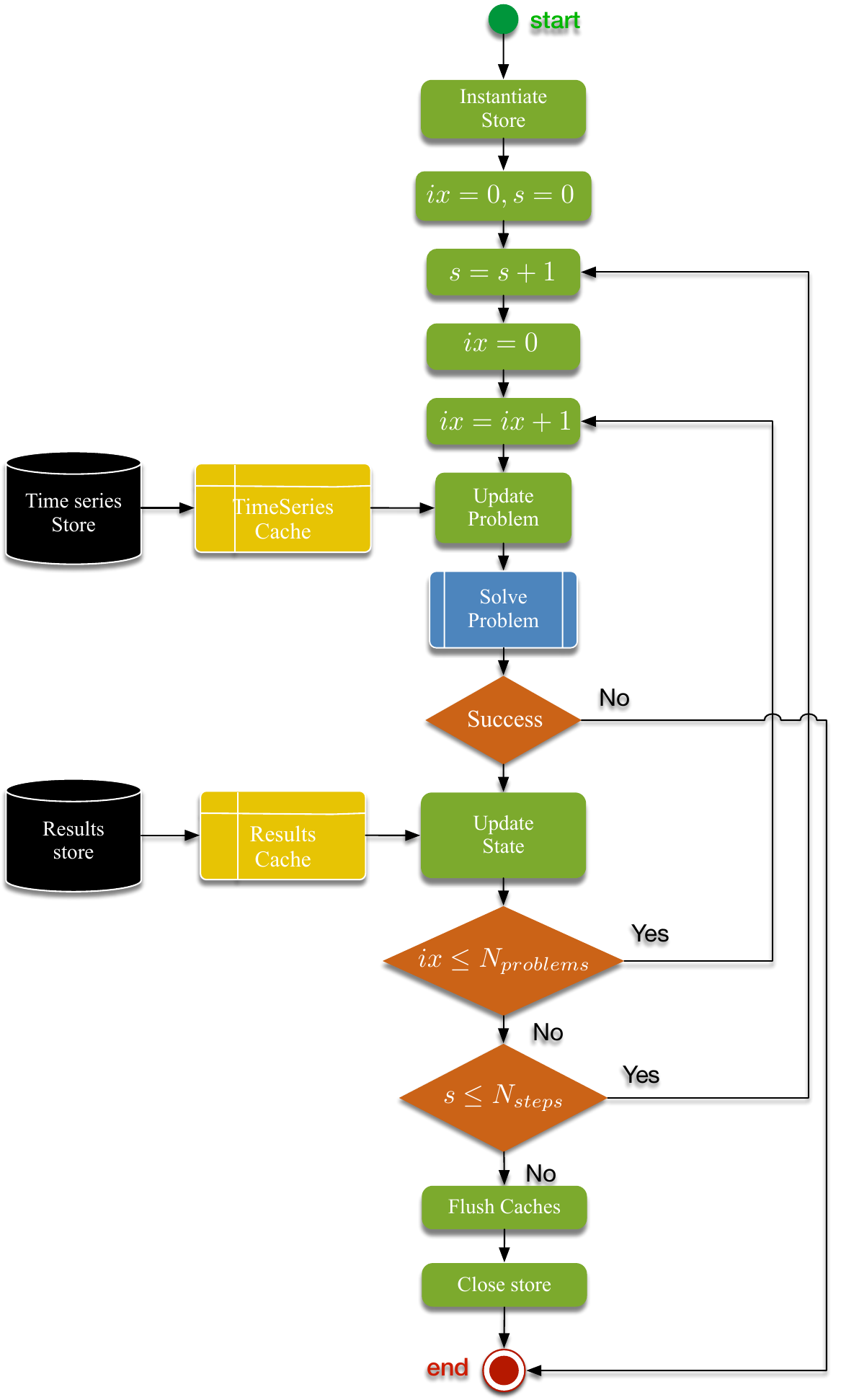}
    }
    \caption{Simulation Execution Process Flow Chart.}
    \label{fig:sim_loop}
\end{figure}

The execution flow of the simulation is shown in Fig. \ref{fig:sim_loop}. The first operation in the execution of a simulation is the instantiation of the \emph{results store}. We developed two implementations of a store: in-memory and HDF5. The in-memory implementation is sufficient for situations where the total data size is smaller than system memory and the analyst doesn't need to access the data beyond one working session. For most cases, however, users will want to store outputs for the long term and so the data must be written to files. We chose the HDF5 format over alternatives like CSV for these reasons:
\begin{itemize}
    \item Supports multidimensional arrays.
    \item Supports inline compression.
    \item Its hierarchical nature allows storing data in a self-describing way.
\end{itemize}

Further, to address the aforementioned shortcomings of write-to-disk approaches, \texttt{PowerSimulations.jl} features a caching layer above the HDF5 file in order to 1) Ensure that all file system writes are at least 1 MiB in size and avoid excessive disk write operations and Provide a read cache for frequently-read outputs.

The instantiation of the \texttt{Sequence} pre-calculates the problem solve execution order in each step. For a simulation of $N_{steps}$ at every simulation step, $s$ a total of $N_{problems}$ are solved sequentially as shown in Fig. \ref{fig:sequence_example1}.

Before execution, the models parameters and initial conditions need to be updated to match the simulation timestamp. The time-series updating uses the caching mechanism from \texttt{PowerSystems.jl} to reduce latency and memory overhead. The feedforward parameters read the latest information from the system state. Once the parameters are updated, the optimizer is executed. After each optimizer reaches successful termination conditions, the results are written into the store and the system states gets updated based on the solution of the optimization model results and the cache of solutions from other models.


%



\bibliographystyle{IEEEtran}
\bibliography{references}

\begin{thebibliography}{10}
\providecommand{\url}[1]{#1}
\csname url@samestyle\endcsname
\providecommand{\newblock}{\relax}
\providecommand{\bibinfo}[2]{#2}
\providecommand{\BIBentrySTDinterwordspacing}{\spaceskip=0pt\relax}
\providecommand{\BIBentryALTinterwordstretchfactor}{4}
\providecommand{\BIBentryALTinterwordspacing}{\spaceskip=\fontdimen2\font plus
\BIBentryALTinterwordstretchfactor\fontdimen3\font minus
  \fontdimen4\font\relax}
\providecommand{\BIBforeignlanguage}[2]{{%
\expandafter\ifx\csname l@#1\endcsname\relax
\typeout{** WARNING: IEEEtran.bst: No hyphenation pattern has been}%
\typeout{** loaded for the language `#1'. Using the pattern for}%
\typeout{** the default language instead.}%
\else
\language=\csname l@#1\endcsname
\fi
#2}}
\providecommand{\BIBdecl}{\relax}
\BIBdecl

\bibitem{KROPOSKI2017}
B.~Kroposki, ``Integrating high levels of variable renewable energy into
  electric power systems,'' \emph{Journal of Modern Power Systems and Clean
  Energy}, vol.~5, no.~6, pp. 831--837, 11 2017.

\bibitem{pfenninger2018opening}
S.~Pfenninger, L.~Hirth, I.~Schlecht, E.~Schmid, F.~Wiese, T.~Brown, C.~Davis,
  M.~Gidden, H.~Heinrichs, C.~Heuberger \emph{et~al.}, ``Opening the black box
  of energy modelling: Strategies and lessons learned,'' \emph{Energy Strategy
  Reviews}, vol.~19, pp. 63--71, 2018.

\bibitem{decarolis_case_2012}
J.~F. DeCarolis, K.~Hunter, and S.~Sreepathi, ``The case for repeatable
  analysis with energy economy optimization models,'' \emph{Energy Economics},
  vol.~34, no.~6, pp. 1845--1853, Nov. 2012.

\bibitem{LARA2020106680}
\BIBentryALTinterwordspacing
J.~D. Lara, J.~T. Lee, D.~S. Callaway, and B.-M. Hodge, ``Computational
  experiment design for operations model simulation,'' \emph{Electric Power
  Systems Research}, vol. 189, p. 106680, 2020. [Online]. Available:
  \url{http://www.sciencedirect.com/science/article/pii/S0378779620304831}
\BIBentrySTDinterwordspacing

\bibitem{plexos}
C.~I. Nweke, F.~Leanez, G.~R. Drayton, and M.~Kolhe, ``Benefits of
  chronological optimization in capacity planning for electricity markets,'' in
  \emph{2012 IEEE International Conference on Power System Technology
  (POWERCON)}, 2012, pp. 1--6.

\bibitem{wang2016quantifying}
Q.~Wang, C.~B. Martinez-Anido, H.~Wu, A.~R. Florita, and B.-M. Hodge,
  ``Quantifying the economic and grid reliability impacts of improved wind
  power forecasting,'' \emph{IEEE Trans. on Sustainable Energy}, vol.~7, no.~4,
  pp. 1525--1537, 2016.

\bibitem{PyPSA}
\BIBentryALTinterwordspacing
T.~Brown, J.~H\"orsch, and D.~Schlachtberger, ``{PyPSA: Python for Power System
  Analysis},'' \emph{Journal of Open Research Software}, vol.~6, no.~4, 2018.
  [Online]. Available: \url{https://doi.org/10.5334/jors.188}
\BIBentrySTDinterwordspacing

\bibitem{ela_flexible_2011}
E.~Ela, M.~Milligan, and M.~O'Malley, ``A flexible power system operations
  simulation model for assessing wind integration,'' in \emph{2011 {IEEE}
  {Power} and {Energy} {Society} {General} {Meeting}}, Jul. 2011, pp. 1--8.

\bibitem{coffrin2018powermodels}
C.~Coffrin, R.~Bent, K.~Sundar, Y.~Ng, and M.~Lubin, ``Powermodels. j1: An
  open-source framework for exploring power flow formulations,'' in \emph{2018
  Power Systems Computation Conference (PSCC)}.\hskip 1em plus 0.5em minus
  0.4em\relax IEEE, 2018, pp. 1--8.

\bibitem{dorris2021choice}
G.~Dorris and D.~Millar, ``Making the right resource choice requires making the
  right model choice,'' \emph{{NRRI} Insights}, pp. 1--7, 2021.

\bibitem{pechman1993model}
C.~Pechman, ``Model-limited choice and the determination of the need for
  generation capacity,'' in \emph{Regulating Power}.\hskip 1em plus 0.5em minus
  0.4em\relax Springer, 1993, pp. 77--97.

\bibitem{panciatici2014advanced}
P.~Panciatici, M.~C. Campi, S.~Garatti, S.~H. Low, D.~K. Molzahn, A.~X. Sun,
  and L.~Wehenkel, ``Advanced optimization methods for power systems,'' in
  \emph{2014 Power Systems Computation Conference}.\hskip 1em plus 0.5em minus
  0.4em\relax IEEE, 2014, pp. 1--18.

\bibitem{DunningHuchetteLubin2017}
I.~Dunning, J.~Huchette, and M.~Lubin, ``Jump: A modeling language for
  mathematical optimization,'' \emph{SIAM Review}, vol.~59, no.~2, pp.
  295--320, 2017.

\bibitem{legat2021mathoptinterface}
B.~Legat, O.~Dowson, J.~Dias~Garcia, and M.~Lubin, ``{MathOptInterface}: a data
  structure for mathematical optimization problems,'' \emph{INFORMS Journal on
  Computing}, 2021.

\bibitem{bezansonJulia}
J.~Bezanson, A.~Edelman, S.~Karpinski, and V.~B. Shah, ``Julia: {{A Fresh
  Approach}} to {{Numerical Computing}},'' \emph{SIAM Review}, vol.~59, no.~1,
  pp. 65--98, 2017.

\bibitem{lara2021powersystems}
J.~D. Lara, C.~Barrows, D.~Thom, D.~Krishnamurthy, and D.~Callaway,
  ``Powersystems. jl—a power system data management package for large scale
  modeling,'' \emph{SoftwareX}, vol.~15, p. 100747, 2021.

\end{thebibliography}

%




\end{document}